\documentclass[fleqn,usenatbib]{mnras}
\usepackage{newtxtext,newtxmath}
\usepackage[T1]{fontenc}
\DeclareRobustCommand{\VAN}[3]{#2}
\let\VANthebibliography\thebibliography
\def\thebibliography{\DeclareRobustCommand{\VAN}[3]{##3}\VANthebibliography}

\usepackage{graphicx}	
\usepackage{amsmath}	
\usepackage{float}
\usepackage{bm}

\title[Dispersion measure variations from scintillation]{Determining electron column density fluctuations in a dominant scattering region using pulsar scintillation}

\author[D. J. Reardon et al.]{
Daniel J. Reardon,$^{1,2}$\thanks{E-mail: dreardon@swin.edu.au}
William A. Coles,$^{3}$
\\
$^{1}$Centre for Astrophysics and Supercomputing, Swinburne University of Technology, P.O. Box 218, Hawthorn, Victoria 3122, Australia\\
$^{2}$Australia Research Council Centre for Excellence for Gravitational Wave Discovery (OzGrav)\\
$^{3}$Electrical and Computer Engineering, University of California at San Diego, La Jolla, California, U.S.A.
}
 
\date{Accepted 2023 March 28. Received 2023 March 06; in original form 2022 August 23}

\pubyear{2020}

\begin{document}
\label{firstpage}
\pagerange{\pageref{firstpage}--\pageref{lastpage}}
\maketitle

\begin{abstract}

Density fluctuations in the ionised interstellar medium have a profound effect on radio pulsar observations, through angular scattering, intensity scintillations, and small changes in time delays from dispersion. Here we show that it is possible to recover the variations in dispersive delays that originate from a dominant scattering region using measurements of the dynamic spectrum of intensity scintillations, provided that the pulsar velocity and scattering region location are known. We provide a theoretical framework for the technique, which involves estimating the phase gradient from the dynamic spectra and integrating that gradient to obtain phase variations. It can be used to search for ``extreme scattering events" (ESEs) in pulsars for which precision dispersion delay measurements are not otherwise possible, or to separate true dispersion variations from apparent variability caused by frequency-dependent pulse shape changes. We demonstrate that it works in practice by recovering an ESE in PSR J1603$-$7202, which is known from precision dispersion delay measurements from pulsar timing. For this pulsar, we find that the phase gradients also track the long-term variations in electron column density observed by pulsar timing, indicating that the column density variations and the scattering are dominated by the same thin scattering screen. We identify a sudden increase in the scintillation strength and magnitude of phase gradients over $\sim$days in 2010, indicating a compact structure. A decrease in the electron density in 2012 was associated with persistent phase gradients and preceded a period of decreased scintillation strength and an absence of scintillation arcs.

\end{abstract}




\section{Introduction}
The ionised interstellar medium (IISM) is a turbulent plasma that disperses and scatters radio-frequency radiation. The frequency-dependent dispersion by this plasma results from the total electron column density along the line-of-sight (LOS), which is referred to as the ``dispersion measure" (DM). Pulsar observations must be corrected for the pulse delays resulting from DM, so that the pulse profile can be integrated over a useful bandwidth. However, precision timing observations, such as those made with pulsar timing arrays to search for gravitational waves, must also be corrected for sub-microsecond errors caused by small time variations in DM. These variations, $\Delta {\rm DM}(t)$, are typically of order 1:10$^4$, and can be measured to a precision of order 1:10$^5$ at cm wavelengths if large fractional bandwidths are available \citep[e.g.][]{Keith+13, Jones+17, Donner+20}. It is somewhat easier to measure $\Delta {\rm DM}(t)$ at meter wavelengths, but it can be complicated by scattering delays, which scale as $\sim \lambda^{4}$ and can become dominant at longer wavelengths.

$\Delta {\rm DM}(t)$ is generally measured by pulsar timing arrays at their regular cadence $\sim$3 weeks, but observations are often averaged to timescales of several months to obtain the necessary precision. Thus the minimum detectable spatial scale is of AU order in current datasets. 
 
 Intensity scintillation and $\Delta$DM(t) variations in the IISM are spatial patterns that drift across the observer. The scintillation shows both diffractive and refractive spatial scales, which have corresponding time scales at cm wavelengths of order 10s of minutes and 10s of hours respectively \citep{Rickett90}. Therefore $\Delta {\rm DM}(t)$ measurements, with time scales of weeks, probe the interstellar turbulence on much larger spatial scales than the scattering observations. Comparing the power at these widely different scales can provide a precise estimate of the spectral exponent of the turbulence. These larger scales also often exhibit non-stationary behavior such as ``extreme scattering events" \citep[ESE;][]{Fiedler+87, Coles+15, Stinebring+22}, which are still not understood in spite of 30 years of work.

Although $\Delta {\rm DM}(t)$ observations are intrinsically interesting and important in correcting precision timing observations, they are not always possible. Here we propose a technique for estimating the $\Delta {\rm DM}(t)$ for pulsars in which diffractive intensity scintillation as a function of time and observing frequency can be measured in a dynamic spectrum. If the scattered image of the pulsar is centered on its true location, then the autocovariance function (ACF) of the dynamic spectrum will be symmetric. However, if there is a mean phase gradient across the scattering disc the scattered image will be displaced as a function of frequency and the ACF will become skewed \citep{Rickett+14}. This skewness is caused by the component of the gradient in the direction of the velocity. It is easily seen in the ACF, but it is also visible in the two-dimensional Fourier transform of the ACF, the ``secondary spectrum." If a parabolic arc is present, the phase gradient will displace the apex of that arc from the origin and produce an asymmetric distribution of power \citep{Cordes+06}.

The skewness is observed in the form of a temporal shear, which must be converted to a spatial shear using the effective velocity of the line of sight through the scattering medium, $\mathbf{V}_\text{eff}$. This requires that the scattering medium be at a known compact location on the line of sight, so $\mathbf{V}_\text{eff}$ can be calculated from the pulsar velocity and the Earth velocities. The gradient of the phase so determined is in the direction of $\mathbf{V}_\text{eff}$. This gradient can then be used to reconstruct the phase as a path integral simply by summing the gradient observations. The temporal summation must also be converted to a spatial integral so again multiplication by $\mathbf{V}_\text{eff}$ is required. The total phase can then be directly converted to an estimate of $\Delta$DM$(t)$.

The technique is very sensitive and can be used on young pulsars which are not as rotationally stable as millisecond pulsars, but it requires that the scattering be dominated by a compact region at a known location and $\mathbf{V}_\text{eff}$ be known. It is particularly effective at detecting ESEs which are, by definition, compact and often show parabolic arcs which make estimation of $\mathbf{V}_\text{eff}$ more accurate.



We first validate the technique and test the code using a simulation that represents ideal conditions. We then apply this technique to a millisecond pulsar observed by the Parkes Pulsar Timing Array \citep[PPTA;][]{Manchester+13}, PSR~J1603$-$7202. This pulsar is of interest because it showed an anomaly in the measured $\Delta {\rm DM}(t)$ from pulsar timing, which was analysed and shown to be an ESE by \citet{Coles+15}. Since the ESE is dense and turbulent enough to dominate the scattering and the $\Delta {\rm DM}(t)$ variations, it is a good candidate for validation of this new method. 

In Section \ref{sec:theory} we present the theoretical background for the technique. In Section \ref{sec:methods} we describe the phase gradient measurement methods and validate the DM recovery technique using a simulation. We describe the observations of the millisecond pulsar PSR~J1603$-$7202 in Section \ref{sec:observations}. The results from this pulsar are presented in Section \ref{sec:results}, with further discussion and speculation on future applications in Section \ref{sec:discussion}. Our conclusions are in Section \ref{sec:conclusions}.
\begin{figure}
\centering
\includegraphics[width=.5\textwidth]{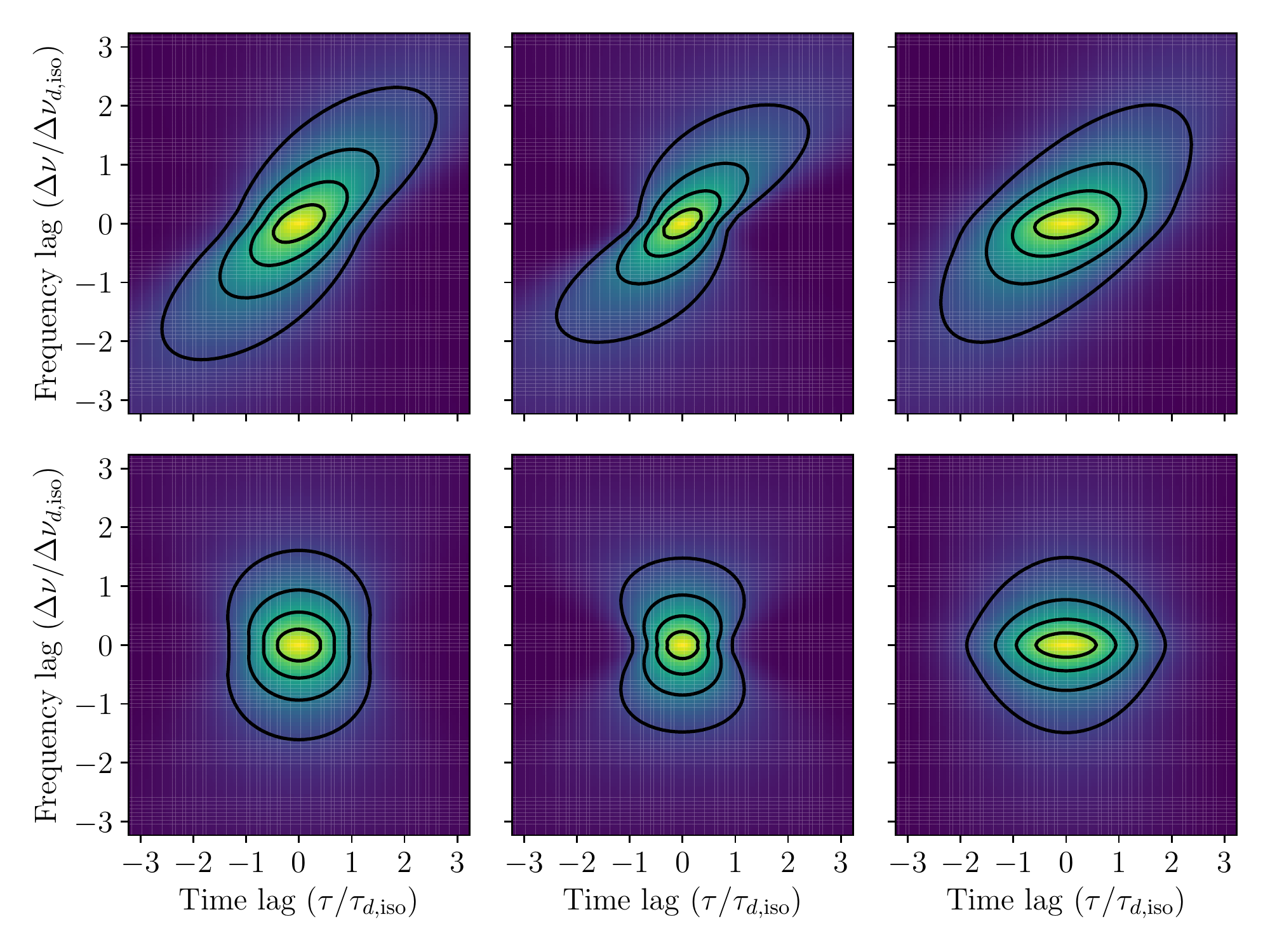}
\caption{The effects of phase gradients (with magnitude 0.4$/s_d$) and anisotropy on the autocovariance of intensity scintillations. The left column is isotropic, the right two columns have axial ratio of 2 with the center oriented parallel to the velocity and the right perpendicular. The top row shows a phase gradient parallel to the velocity and for the bottom row it is perpendicular.
The axes have been normalised using the characteristic scintillation scales of isotropic scattering assuming no phase gradient present, $\Delta\nu_{d,\rm{iso}}$ and $\tau_{d,\rm{iso}}$.}
\label{fig:acf2}
\end{figure}

\section{Theory}
\label{sec:theory}
\begin{figure*}
\centerline{\includegraphics[width=\textwidth, trim= 0 0 0 0 , clip]{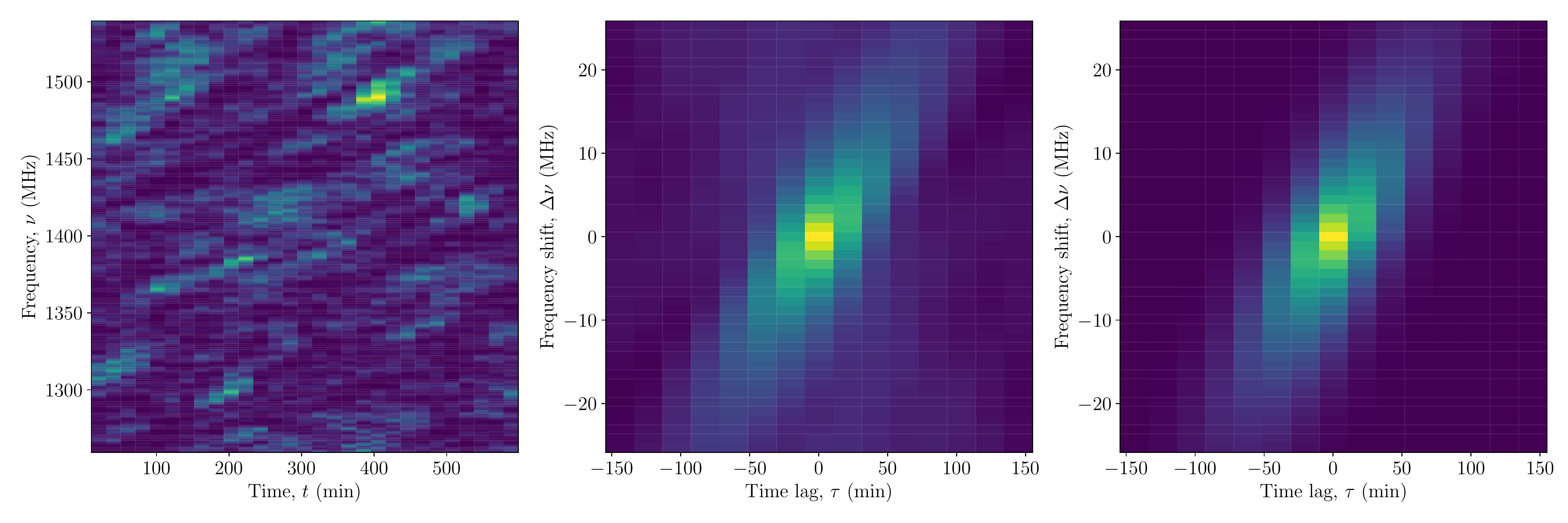}}
\caption{The dynamic spectrum (left), ACF (middle), and model ACF of Equation \ref{eqn:approx} (right) for the first observation drawn from the simulation. The measured shear in the ACF is $\mu = 2.65 \pm 0.04\,$min$\,$MHz$^{-1}$.}
\label{fig:dynspec2}
\end{figure*}
\subsection{Intensity Scintillation}
The theory of intensity scintillation is well-developed and we will only summarize the results here \citep[see][]{Rickett77, Rickett90, Narayan92, Cordes+06}. The underlying phenomenon is angular scattering caused by refractive index fluctuations in the scattering medium. These cause phase fluctuations in the radio wave radiated by the pulsar. In the interstellar plasma, these fluctuations are described by
\begin{equation}
\phi(\mathbf{x}) = -\lambda r_e \int_0^D n_e(\mathbf{x} z/D, z) dz.
\end{equation}
Here $\mathbf{x}$ is the position transverse to the LOS, $n_e$ is the electron density, $z$ is the distance along the LOS from the pulsar, the observer is located at $z=D$, $\lambda$ is the wavelength, and $r_e$ is the classical electron radius. The dispersion measure (DM) is simply the electron column density.

If the scattering medium has stationary gaussian differences then the angular spectrum is described by the phase structure function 
\begin{equation}
D_\phi (\boldsymbol{\sigma}) = \langle(\phi(\mathbf{x}) - \phi(\mathbf{x}+\boldsymbol{\sigma}))^2\rangle.
\end{equation}
The angle brackets denote ensemble average and $\boldsymbol{\sigma}$ is a spatial offset from position $\mathbf{x}$, with both vectors transverse to the LOS in the observer's plane. The width of the angular spectrum is defined with $\theta_0 = 1/(ks_d)$, where $D_\phi(s_d) = 1$ and
$k = 2 \pi / \lambda$. The phase structure function generally has a power law form, $D_\phi (\boldsymbol{\sigma}) = (\boldsymbol{\sigma}/s_d)^{\alpha}$, where $s_d$ is the coherence spatial scale. The exponent $\alpha = 5/3$ corresponds to Kolmogorov turbulence. The autocovariance of the electric field $C_E(\boldsymbol{\sigma})$ is
\begin{equation}
C_E(\boldsymbol{\sigma}) = \exp(-0.5 D_\phi (\boldsymbol{\sigma})).
\end{equation}
 Thus the Fourier transform of $C_E(\boldsymbol{\sigma})$, which is the angular spectrum of plane waves, or the brightness distribution, is also completely determined by $D_\phi (\boldsymbol{\sigma})$.

In the very common case where most of the scattering takes place in a ``thin screen", the bulk of the radiation received by the observer passes through the ``scattering disc" which has radius $s_r = \theta_0 D_{\rm eff}$, where $D_{\rm eff}$ is the effective distance from the observer to the scattering region. The angular spectrum broadens pulses to a scattering timescale $t_s = \theta_0 ^2 D_{\rm{eff}} / 2 c$. 

The intensity scintillations, which are caused by interference between the scattered plane waves, are then correlated over a bandwidth $\Delta\nu_{d} = 1/2 \pi t_s$. When this bandwidth is small compared with the observing bandwidth, the scintillation is said to be ``strong" and the autocorrelation of intensity $C_I(\boldsymbol{\sigma}) = |C_E(\boldsymbol{\sigma})|^2$. The autocorrelation in time is simply $C_I(\boldsymbol{\tau}) = C_I(\boldsymbol{\sigma} =\mathbf{V}_{\rm eff} \tau)$. Pulsar observations are almost always observed in strong scintillation. In this case there are also slower refractive intensity scintillations on the spatial scale of the scattering disc. This scale is usually much longer than the observation duration and is seldom measured for pulsars.

Intensity scintillation is observed as a time variation but it is actually due to a spatial variation convected across the LOS with effective velocity
\begin{equation}
\mathbf{V}_{\rm eff} = s\mathbf{V}_{\rm E} + (1-s)\mathbf{V}_{\rm p} - \mathbf{V}_{\rm IISM}.
\end{equation}
Here the velocities of the observer, the IISM, and the pulsar are $\mathbf{V}_{\rm E}$, $\mathbf{V}_{\rm IISM}$ and $\mathbf{V}_{\rm p}$ respectively, and the fractional distance from the pulsar to the thin screen is $s$, with the observer at $s=1$.

The transverse fluctuations in $\phi$ may be anisotropic, and if so the timescale will depend on the direction of the velocity as well as its magnitude. Observers usually measure the timescale $\tau_{d}$ and bandwidth $\Delta\nu_{d}$ from a two dimensional autocovariance of the dynamic spectrum $C_I (\tau,\Delta\nu)$. This can be modeled analytically and provides information on the anisotropy \citet{Rickett+14}. The autocovariance in frequency $C_I (\Delta\nu)$ is more complex than $C_I (\tau)$, but an analytic form is available for a thin screen in Equations A1 and A2\footnote{There is an error in Equation A2. In the two places where $\nu_m$ (centre frequency) appears, it should be replaced by $2\pi\nu_m$.} of \citet{Rickett+14}. 

It is often the case that the turbulence is somewhat anisotropic and occasionally it is very anisotropic \citep{Brisken+10}. In such cases $D_\phi(\boldsymbol{\sigma})$ becomes a quadratic form and a 2-D analysis is required. The evaluation of $C_I (\tau,\Delta\nu)$ is shown in Figure 12 of \citet{Rickett+14}, demonstrating the effect of different anisotropy axial ratios $A_r$, with $\mathbf{V_{\rm eff}}$ parallel to the major axis.

\subsection{Phase Gradients}
If there is a gradient of $\phi$ over the scattering disc, the apparent position of the pulsar will be displaced by an angle $\mathbf{\theta}_p = \nabla{\boldsymbol{\phi}}/k$, where $\nabla{\boldsymbol{\phi}}$ is the mean of the gradient. This will displace the diffraction pattern by a distance $\boldsymbol{\sigma}_p = Ds(1-s)\mathbf{\theta}_p$. Since $\mathbf{\theta}_p \propto \lambda^2$, this displacement varies with frequency. The phase gradient therefore produces a chromatic aberration of the pulsar's scattered image.

If the gradient has a component in the direction of $\mathbf{V}_{\rm eff}$, this causes the tilted bands often observed in pulsar dynamic spectra. This skews the ACF as shown in the top row of Figure \ref{fig:acf2}. Its corresponding effect on secondary spectra (asymmetries) is described in \citet{Cordes+06}. A phase gradient perpendicular to $\mathbf{V_{\rm eff}}$ narrows the ACF symmetrically in $\Delta\nu$, which is shown in the bottom row of Figure \ref{fig:acf2}. As this does not skew the ACF it is difficult to distinguish from stronger scintillation.

For small bandwidths the angular displacement can be linearised, resulting in Equation A6 of \citet{Rickett+14}
\begin{equation}
C_I  ( \tau, \Delta\nu ) = C_I ( \boldsymbol{\sigma} =\mathbf{V}_{\rm eff}\tau  - 2 \boldsymbol{\sigma}_p (\Delta \nu / \nu), \Delta \nu).
\end{equation}A cut through the 2D ACF at fixed $\Delta\nu$ will therefore peak at 
\begin{equation}
\label{eqn:tpk}
\tau_{\rm pk} = \frac{2 Ds(1-s)}{k V_{\rm eff}}\left(\frac{\Delta\nu}{\nu}\right) \nabla\phi_\parallel ,
\end{equation} where $\nabla\phi_\parallel = (\mathbf{V}_{\rm eff}\cdot\nabla\phi) / V_{\rm eff} $. So by measuring $\tau_{\rm pk}$ from the ACF we can determine the phase gradient in the direction of the velocity, $\nabla\phi_\parallel$. What has not been realized earlier, is that by integrating $\nabla\phi_\parallel$ over time, we perform a path integral along the trajectory of the LOS through the IISM, and recover temporal variations in the total phase,
\begin{equation}
\label{eqn:int}
\Delta\phi(t) = \oint \nabla\phi_\parallel(t)V_{\rm eff}(t) dt .
\end{equation}
The shear is easily determined, regardless of the shape of the ACF, because it is simply the constant required to make the de-skewed ACF symmetrical. Therefore, the phase gradients can be estimated from the skew, regardless of whether the simple Kolmogorov scattering model used in the previous section accurately describes the data.

It is important to note that the skew does not affect the temporal ACF $C_I(\tau,0)$, so the $\tau_d$ is unchanged. However it does narrow the bandwidth if it is determined from $C_I(0,\Delta\nu)$. Clearly $\Delta\nu_{d}$ should be measured after deskewing the ACF, but this has not been the common practice. Phase gradients perpendicular to the velocity, which do not skew the ACF, do narrow the bandwidth. Observed bandwidth estimates will show more variance than expected if these phase gradients are unmodeled.

\subsection{DM Estimation}

Dispersion measure variations can be recovered as $\Delta\rm{DM} = -\Delta\phi / \lambda r_e$. With $\Delta\phi$ in radians and $\nu$ in GHz, the DM variations in typical units of pc$\,$cm$^{-3}$ are given by 
\begin{equation}
    \Delta {\rm DM} = -3.84 \times 10^{-8}\nu\Delta\phi .
\end{equation}
This $\Delta {\rm DM}$ is determined from our observations by integrating the estimates of $\nabla\phi_\parallel$ over time according to the path integral in Equation \ref{eqn:int}.

Both $\nabla\phi_\parallel$ and $\Delta \phi$ require $V_{\rm eff}$, so the location of the scattering medium, which is seldom known accurately, is a primary source of error \citep{Rickett+14, Reardon+19}. In this respect it is very helpful if a scintillation arc is observed as it will locate the scattering region even if the scattering is anisotropic \citep[e.g.][]{Reardon+20, Walker+22}.

The purpose of this paper is to bring attention to the fact that an analysis of small-scale diffractive scintillations can allow us to recover $\Delta {\rm DM}(t)$ on much larger scales, so we have neglected the anisotropy of the phase structure function and consider only the simpler case of isotropic scattering.  While a thin screen can often dominate the scattering, which is proportional to a path integral over $n_e^2$, it may not dominate the dispersion, which is a path integral over $n_e$. Hereafter we use the notation $\Delta {\rm DM}_\phi$ to represent the transverse DM variations derived from $\nabla\phi_\parallel$ using scintillation. It may differ from the total $\Delta {\rm DM}$ if there is a significant variation caused by the radial motion of the pulsar in a high density environment, or if the screen density variations do not dominate the LOS.

\section{Methods}
\label{sec:methods}
As derived in the previous Section, a phase gradient in the direction of the LOS velocity, $\nabla\phi_\parallel$, causes chromatic aberration which is observed as a shear to the time-frequency ACF, $C_I  ( \tau, \Delta\nu )$. We measure this shear, $\mu$ directly from $C_I  ( \tau, \Delta\nu )$ as the gradient from the relation $\tau_{\rm pk} = \mu \Delta\nu$. From Equation \ref{eqn:tpk}, we convert this to $\nabla\phi_\parallel$ using a model of $V_{\rm eff}$, $D$, and $s$
\begin{equation}
\label{eqn:nablaphi}
    \nabla\phi_\parallel = \frac{\mu k \nu V_{\rm eff}}{2Ds(1-s)}
\end{equation}
For discrete measurements of this gradient, such as in real pulsar observations, the path integral of Equation \ref{eqn:int} can be approximated with weighted sum
\begin{equation}
\label{eqn:phisum}
    \Delta\phi (t) \approx \sum^{N(t) - 1}_{i=0} \nabla\phi_{\parallel}(t_i)V_{\rm eff}(t_i)(t_{i+1} - t_i),
\end{equation}
where $t_i$ is the epoch of the ith observation, and $N(t)$ is the number of observations between $t$ and $t_0$. The time series of $\Delta\phi (t)$ is therefore a weighted cumulative sum of $\nabla\phi_{\parallel}$. In performing this cumulative sum, the random measurement errors accumulate into correlated noise with a power spectrum that scales as $f_t^{-2}$, for conjugate time $f_t$. We can quantify this noise and the uncertainty on $\Delta\phi(t)$ by simulating our random measurements and performing the weighted sum for each simulated dataset.

\subsection{Measuring the ACF shear}
\label{sec:tilts}

Here we describe two methods for measuring shear/tilt parameter $\mu$ from the time-frequency ACFs.

In the first, we determine $\tau_{\rm pk}$ at multiple $\Delta\nu$ values near the core of the ACF ($\Delta\nu < \Delta\nu_d$), and then simply fit a straight line to these measurements, $\tau_{\rm pk} = \mu \Delta\nu$. The $\tau_{\rm pk}$ measurements are made to a fraction of $\delta t$ in precision, by fitting an inverse parabola to the three samples across the peak of the ACF at each $\Delta\nu$.

In the second method, we approximate the shape of the ACF in two-dimensions with a model that is fast to compute, unlike the full model of \citet{Rickett+14} discussed in Section \ref{sec:theory}. The model has an exponential form along $C_m(0, \Delta\nu)$, and a variable form along $C_m(\tau, 0)$,
\begin{align}
\label{eqn:approx}
C_m(\tau, \Delta\nu) &= A \exp \left(- \left|\left(\frac{\tau - \mu \Delta\nu}{\tau_d}\right)^{3\alpha/2} \left(\frac{\Delta\nu}{\Delta\nu_{d}/\log(2)}\right)^{3/2}\right|^{2/3}\right)\\
C_m(0, 0) &= A + w \nonumber,
\end{align}
where $A$ is the amplitude and $w$ is the white noise level. We use $\alpha = 5/3$ for the Kolmogorov form of $C_m(\tau, 0)$, and following convention $\Delta\nu_d$ is the half-power scale in frequency, while $\tau_d$ is the 1/e scale in time. The model is an approximation that is accurate for near-isotropic scattering. Modelling anisotropic scattering requires the more complex analytical model from \citet{Rickett+14}.

After measuring the shear $\mu$, and its uncertainty using one of these methods, we estimate $\nabla\phi_\parallel$ using Equation \ref{eqn:nablaphi}. The uncertainty is derived from the measurement error, as well as a finite scintle error estimate \citep{Cordes+86} that is added in quadrature. We use Equation \ref{eqn:approx} to model the simulations, as the scattering is isotropic by design. However, we use the first method (a straight line fitted directly to $\tau_{\rm pk}$ measurements) for some observations of PSR~J1603$-$7202 (when Equation \ref{eqn:approx} gives a poor fit), because there is evidence for anisotropy both in the ACFs, and in the spectra \citep{Walker+22}. 
\begin{figure}
\centering
\includegraphics[width=.5\textwidth]{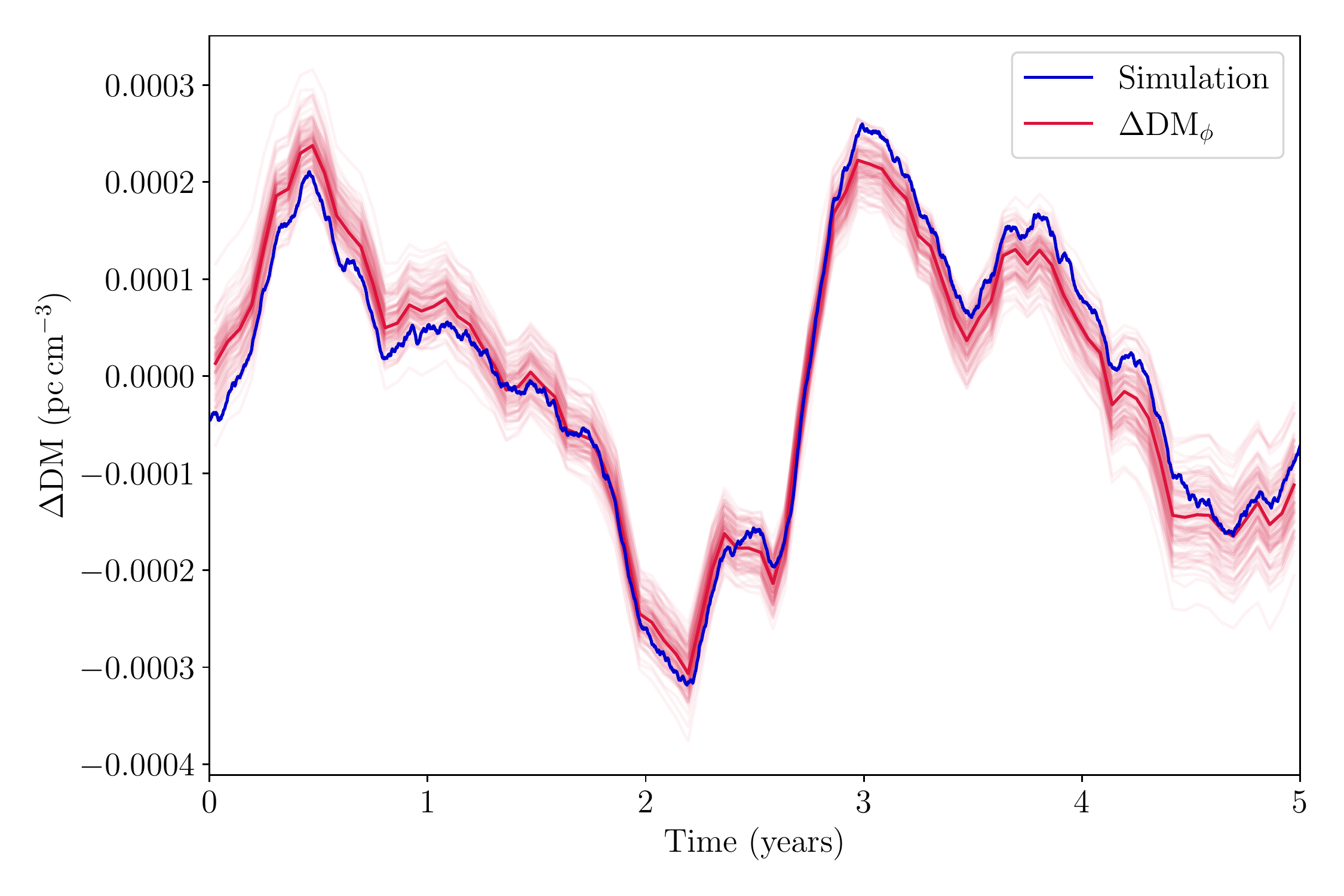}
\caption{The dispersion measure variations from the simulation described in Section \ref{sec:simulation}. The blue line shows the exact $\Delta{\rm DM}$ caused by the phase variations that arise naturally in our simulated scattering screen. The dark solid red line and faint red lines show the recovered $\Delta{\rm DM}_\phi$ and 100 simulations of $\Delta{\rm DM}_\phi$ respectively.}
\label{fig:totalphase}
\end{figure}
\begin{figure*}
\centerline{\includegraphics[width=\textwidth, trim= 0 0 0 0 , clip]{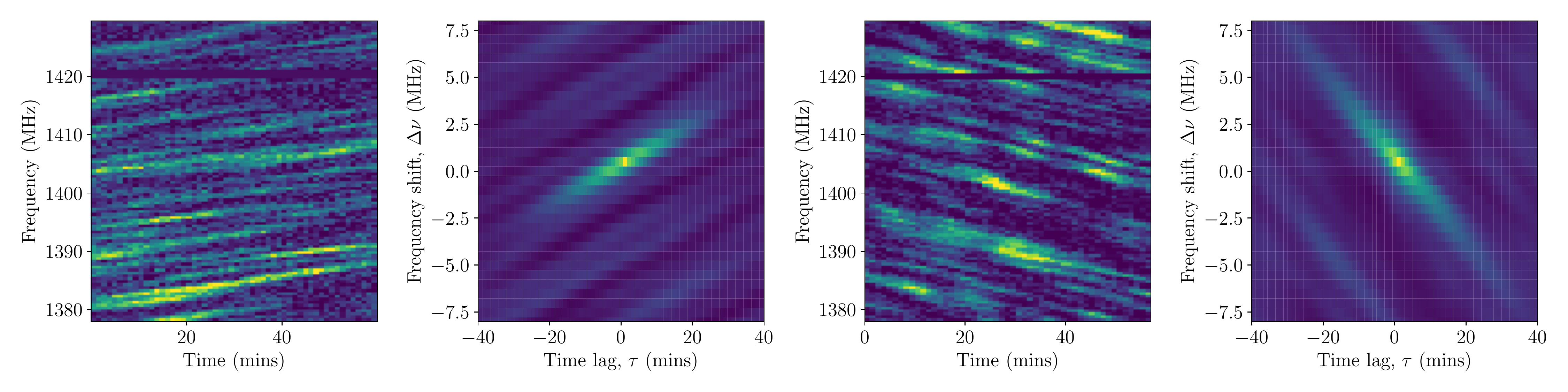}}
\caption{Dynamic spectra and auto-correlation functions of PSR~J1603$-$7202 during observations taken near the start (left two panels; MJD 53832) and end (right two panels; MJD 53966) of the ESE.}
\label{fig:1603_dynspec}
\end{figure*}

\subsection{Validation via simulation}
\label{sec:simulation}
To test the proposed method we used a full electromagnetic simulation following the techniques described in \citet{Coles+10}, which has been reproduced in the \textsc{scintools} package\footnote{\url{https://github.com/danielreardon/scintools}} \citep{Reardon+20}. We produced one continuous $5\,$yr simulated dynamic spectrum, from which the phase variations with time are derived exactly. The combination of simulation and pulsar parameters (e.g. $s_{\rm d}$, $D$, $\nu$, $\mathbf{V}_{\rm eff}$, and the strength of scintillation) were chosen such that the resulting scintillation bandwidth, timescale, and amplitude of DM variations were comparable to many PTA millisecond pulsars.

The sampling characteristics of the dynamic spectrum included a coarse sub-integration time of $\delta t = 20\,$mins, to reduce computation cost of the simulation. The fractional bandwidth was 20\% with 256 channels and a centre frequency of $\nu_c = 1400\,$MHz. Mock observations were sampled from this dynamic spectrum, with a length of 10\,hrs every 20\,days. The observing cadence is comparable to many long-term observing campaigns, but the observing time is longer than real observations. This is to account for the coarse sampling, which increases the measurement uncertainties. In real observations, the measurement uncertainties are be reduced by finer sampling, but the total uncertainty is increased because of additional errors due to the finite number of scintles in the spectrum \citep{Cordes+86}. Our choice of a $10\,$hr observation produces fractional uncertainties comparable to real observations after accounting for finite-scintle errors. In addition, the phase gradient is not observed to evolve significantly across this time.

Using the two-dimensional ACF model in Equation \ref{eqn:approx}, we measured the scintillation scales and ACF shear parameter $\mu$ for each mock observation. The mean and standard deviation for the scintillation scales across the were, $\Delta\nu_{\rm d} = 5.4 \pm 1.9\,$MHz at $\nu = 1400\,$ MHz, and $\tau_{\rm d} = 30 \pm 6\,$mins respectively. While the mean $\tau_{\rm d}$ was only 50\% larger than $\delta t$, we are able to recover this timescale reliably in every observation. The dynamic spectrum, ACF, and model fit for the first observation is shown in Figure \ref{fig:dynspec2}.

The path integral of the phase gradient was computed and compared with the known phase variations produced in the simulation. The comparison is shown in Figure \ref{fig:totalphase}, with the darkest red line showing our measurements, with the fainter lines showing the simulated measurements that represent the uncertainty region. Each $\Delta {\rm DM}(t)$ curve has been set to zero mean. The recovered $\Delta {\rm DM}_\phi(t)$ is seen to randomly deviate from the true value, because of the $f_t^{-2}$ noise induced by performing the cumulative sum. The agreement is remarkably good considering the small duty cycle of the observations. The reason for this is that the $\Delta {\rm DM}(t)$ have a Kolmogorov power law spectrum, and are dominated by the lowest frequencies.  We have tested the process with even smaller duty cycles and confirmed that the match decreases slowly, as one would expect.

This demonstrates that the method can accurately recover DM variations due to a dominant scattering screen, provided that the ACF is modelled accurately (e.g. including any anisotropy) and that the velocity of the LOS through the scattering medium is known.

\section{Observations}
\label{sec:observations}
The millisecond pulsar J1603$-$7202, is regularly observed as part of the Parkes Pulsar Timing Array (PPTA) project, using the 64$\,$m Parkes radio telescope (Murriyang). Here we use the dynamic spectra produced as part of the second data release of the PPTA \citep{Kerr+20}, in the frequency band centred on $\nu=1368\,$MHz, and spanning the MJD range 53000 to 57250. The mean separation between successive observations in our dataset is $\sim 17 \,$days. The ESE in this pulsar lasted $\sim 250\,$days \citep{Coles+15}, and multiple observations at $\nu=1368\,$MHz were recorded during this period.

This pulsar is in a $\sim 6.3\,$day orbit with a white dwarf companion, and the dominant scattering occurs relatively close to the pulsar ($s\sim0.25$), so the influence of the Earth's velocity is small. We can estimate the influence of the pulsar's orbit on $\mathbf{V_{\rm eff}}$ using the screen and orbital geometry inferred from a recent analysis of scintillation arcs by \citet{Walker+22}. While \citet{Walker+22} proposed several potential models for the orbit in their Table 1, we assume the single-epoch, anisotropic model, with longitude of ascending node $\Omega = 327^\circ$, and use this model to perform the weighted sum of Equation \ref{eqn:phisum}. The pulsar distance is assumed to be $D=3.3\,$kpc, and the screen distance $s=0.25$, giving the mean velocity $\langle V_{\rm eff}\rangle = 97\,$km$\,$s$^{-1}$. The scintillation arc study was only sensitive to the component of the screen velocity aligned with the major axis of anisotropy, so we assume the remaining component is small compared with the mean ${V}_{\rm eff}$.

The choice of model for $\mathbf{V}_{\rm eff}$ does affect the recovered $\Delta {\rm DM}_\phi$, however the purpose here is to show primarily that the ESE and other short-timescale structures in the DM variations can be identified, rather than achieving a precise reconstruction. In addition, since the orbital period ($\sim$6.3$\,$days) is smaller than the mean observing cadence ($\sim$17$\,$days), a computation of the phase gradient and orbital velocity generally cannot be extrapolated with precision to the subsequent observation. For this reason, we have simply assumed one velocity model that appears to approximately recover the known $\Delta{\rm DM}(t)$, and do not attempt to use our observations to validate or improve the models of \citet{Walker+22}.

DM variations have been measured during a timing analysis of the PPTA second data release for PSR~J1603$-$7202 \citep{Reardon+21}. The time series of $\Delta {\rm DM}(t)$ was generated using two methods: the piece-wise linear function of \citet{Keith+13} with a 60$\,$day sampling interval (meaning $\Delta {\rm DM}(t)$ is smoothed by a triangle function with a 120$\,$day base), and a powerlaw model of \citet{Lentati+14a} with an additional Gaussian-shaped bump during the ESE. The interpolation measurements are discrete, while the powerlaw model is a stochastic process with time-correlated values and uncertainties. We represent the latter in our figures, as a shaded 68\% confidence region with a solid line through the mean. For our $\Delta{\rm DM}_\phi(t)$, we visualised the correlated uncertainty region by showing 100 simulated datasets produced from our measurements (assuming a Gaussian probability density about their mean).

\section{Results}
\label{sec:results}
Here we describe the results after applying our technique to PSR~J1603$-$7202, which is notabe for showing an extreme scattering event (ESE). We have measured the characteristic scintillation scales $\Delta\nu_{\rm d}$ and $\tau_{\rm d}$, as well as the parallel phase gradients $\nabla\phi_\parallel$ using Equation \ref{eqn:approx}. $\Delta{\rm DM}_\phi (t)$ was recovered using the weighted sum of Equation \ref{eqn:phisum}. The dynamic spectra and ACFs for two observations, early and late, in the ESE are shown in Figure \ref{fig:1603_dynspec}. It is clear that the skewness of the ACF is a dominant characteristic and can be estimated with good precision.

\begin{figure}
\centering
\includegraphics[width=.5\textwidth]{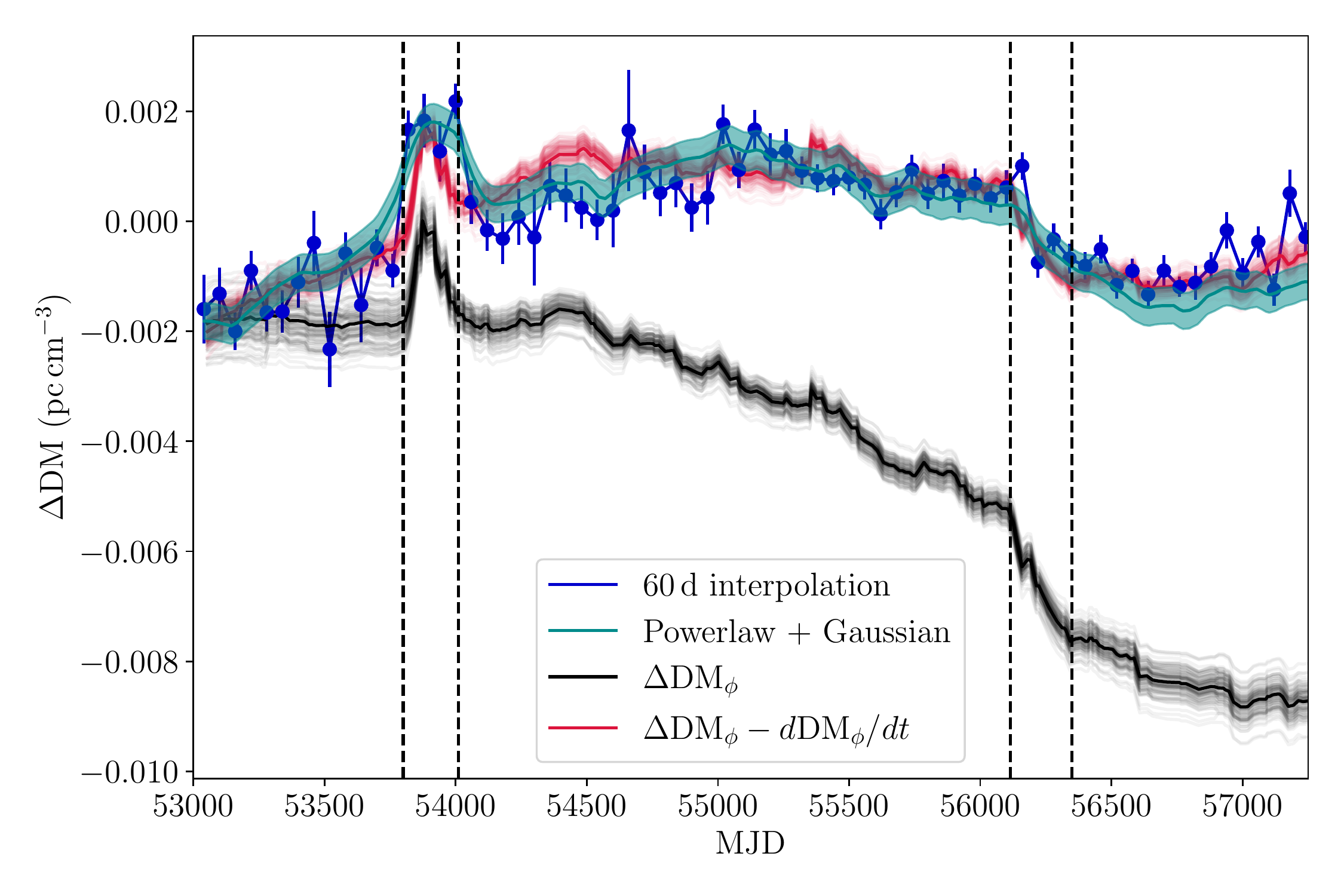}
\caption{DM variations measured from the PPTA-DR2 dataset using a 60-day piece-wise linear function (blue), and as a stochastic process described by a powerlaw, with an additional Gaussian bump during the ESE (teal; shaded region is 68\% confidence interval). The dark and faint black lines show the $\Delta{\rm DM}_{\phi}(t)$ recovered from $\nabla\phi_\parallel$, and 100 simulations of the dataset respectively. The red lines correspond to the black $\Delta{\rm DM}_{\phi}(t)$ lines with a linear trend subtracted. The black $\Delta{\rm DM}_{\phi}(t)$ have been shifted to a mean of $-0.004\,$pc$\,$cm$^{-3}$ for clarity. The left two vertical dashed lines mark the known ESE, and the right two dashed vertical lines mark a period of largely negative phase gradients, a DM decrease, and fewer scintillation arcs \citep{Walker+22}.}
\label{fig:J1603_1}
\end{figure}

The DM variations from pulsar timing (using two methods) and the recovered $\Delta{\rm DM}_\phi$ for PSR~J1603$-$7202 are shown in Figure \ref{fig:J1603_1}. We observe a persistent gradient offset between $\Delta{\rm DM}_{\phi}(t)$ and the measurements from pulsar timing. A gradient was not observed in the simulations, and is larger than expected from the random walk induced during the cumulative sum of $\nabla\phi_\parallel$. We show the original $\Delta{\rm DM}_{\phi}(t)$ (in black) as well as the $\Delta{\rm DM}_{\phi}(t)$ with a the gradient $d{\rm DM}_{\phi}/dt$ removed (in red). Also in Figure \ref{fig:J1603_1} we highlight two regions of interest with dashed lines. The first is the known ESE, which clearly presents in $\Delta{\rm DM}_{\phi}(t)$, proving that the DM$(t)$ in this region is associated with a compact structure that also dominates the scattering. The second marked region shows a steep decline in $\Delta{\rm DM}_{\phi}(t)$, followed immediately by a flattened region, which is also apparent in the pulsar timing measurements. We speculate that this flattening is caused by the end of a dominant scattering region that persisted since the start of the PPTA observations of this pulsar, because it also coincides with the disappearance of scintillation arcs \citep[see Figure 1 of][]{Walker+22}.

\begin{figure*}
\centering
\includegraphics[width=\textwidth, trim= 0 0 0 0, clip]{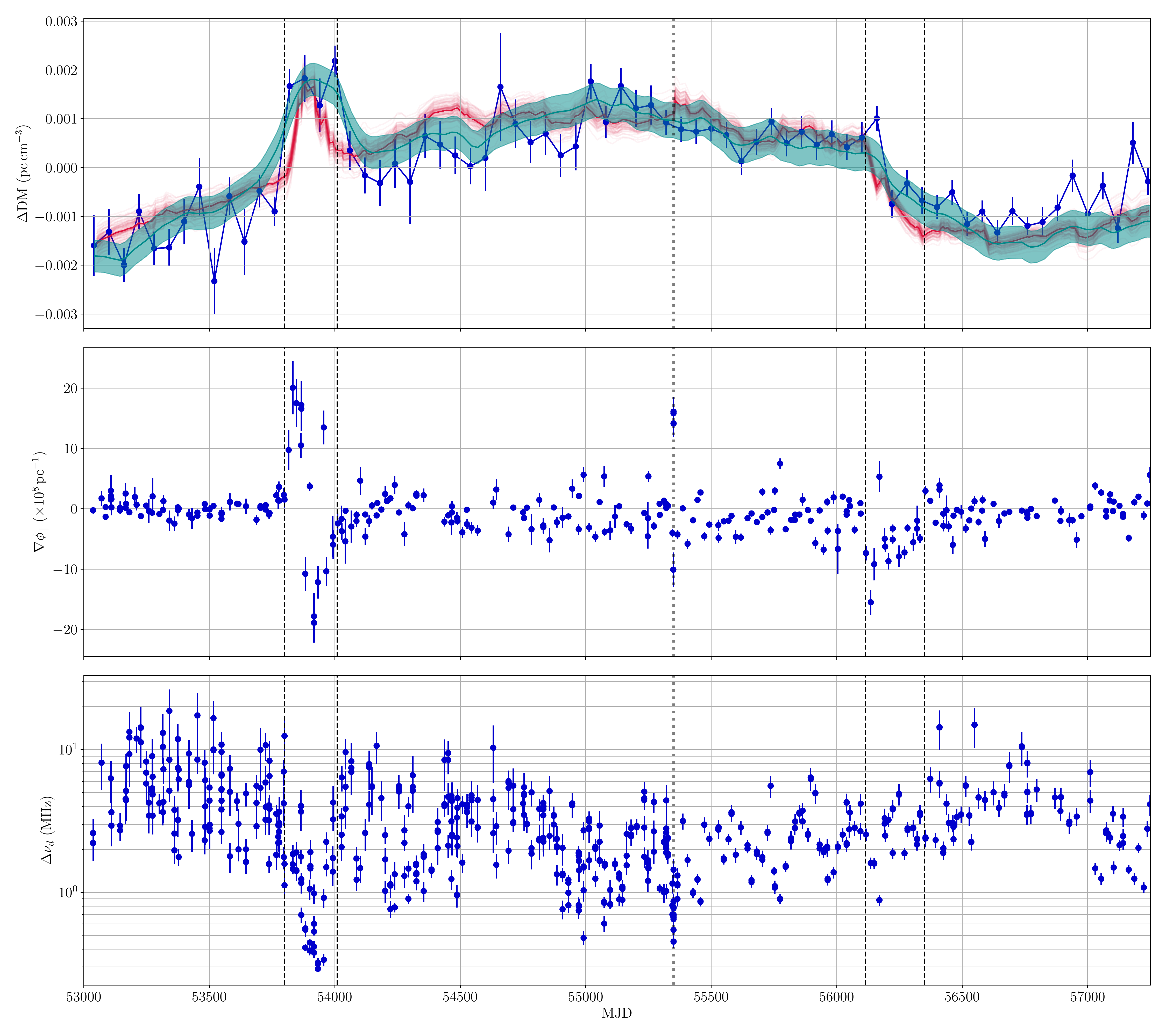}
\caption{Dispersion measure variations (top panel), phase gradients parallel to effective velocity (middle panel), and scintillation bandwidths (bottom panel), for PSR~J1603$-$7202. The top panel shows the DM variations from both timing and scintillation, as in Figure \ref{fig:J1603_1}. However, a gradient offset between $\Delta{\rm DM}_{\phi}(t)$ and the powerlaw data (teal) was fitted and added to $\Delta{\rm DM}_{\phi}(t)$. The two left dashed lines mark the approximate start and end of the ESE. The dotted line marks a potential short-timescale DM increase discussed in-text, and the right two dashed lines mark the speculated end of a dominant scattering region.}
\label{fig:1603}
\end{figure*}

We found that the two-dimensional ACF model in Equation \ref{eqn:approx} was not always appropriate for measuring $\tau_d$ or $\mu$ in this pulsar because at some epochs it shows evidence of anisotropy in the ACF and secondary spectra. However, we find that this model is useful for estimating $\Delta\nu_d$ and gives measurements consistent with other methods \citep[e.g.][]{Reardon+19}. The measured $\Delta\nu_d$, DM variations, and the derived $\nabla\phi_\parallel$ are shown in Figure \ref{fig:1603}. In this Figure, the gradient offset between the powerlaw $\Delta{\rm DM}(t)$ and $\Delta{\rm DM}_{\phi}(t)$ was measured and added to $\Delta{\rm DM}_{\phi}(t)$. 

\label{sec:res_1603}

We measured the gradient offset as $d{\rm DM}/dt = (6.5 \pm 0.7)\times 10^{-4}\,$pc$\,$cm$^{-3}\,$yr$^{-1}$. This $d{\rm DM}/dt$ is attributed to weaknesses in our measurements (e.g. assuming isotropic scattering for the ACF), or our model of $\mathbf{V}_{\rm eff}$. The gradient may also be in part caused by density variations that are not strong enough to dominate the scattering. Since scintillation is only sensitive to the transverse motions of the Earth, IISM, and pulsar, this density gradient could be due in part to the pulsar's radial velocity $V_r$ which contributes to the measured $\Delta{\rm DM}$ derived from pulsar timing. Under this assumption, we can derive constraints on the $V_r$ and density of the IISM local to the pulsar, $n_e$. For example, if $V_r = 100\,$km$\,$s$^{-1}$, then we derive $n_e = 6.4\,$cm$^{-3}$. This is much larger than the average density along the LOS, and two orders of magnitude greater than the predicted density at the pulsar's location from Galactic electron density models \citep{Cordes+02, Yao+17}. The proposed alternative explanations for the presence of a gradient offset are therefore more likely in this case (primarily, errors in the assumed $\mathbf{V}_{\rm eff}$).

\begin{figure*}
\centering
\includegraphics[width=.485\textwidth]{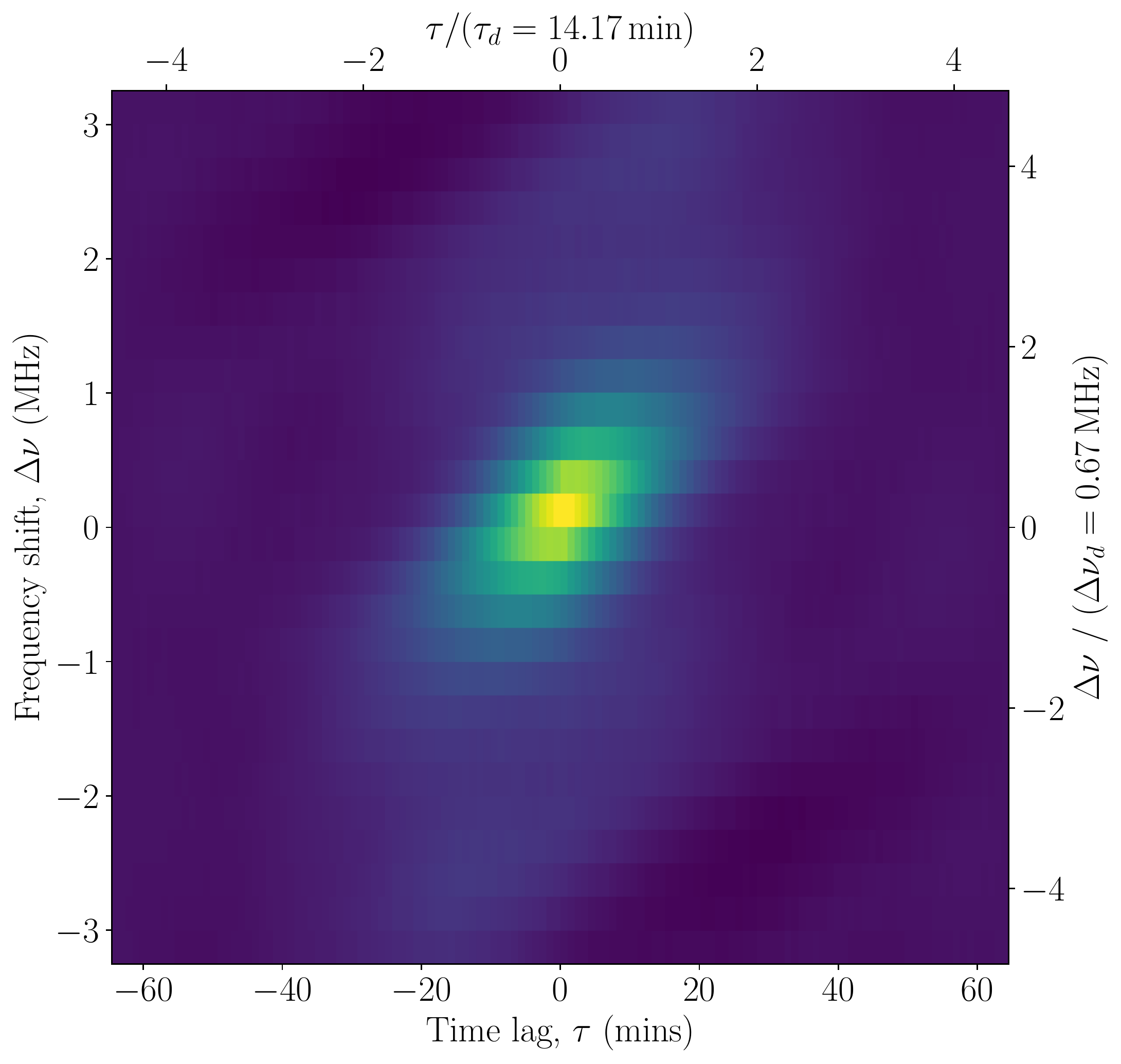} \includegraphics[width=.45\textwidth]{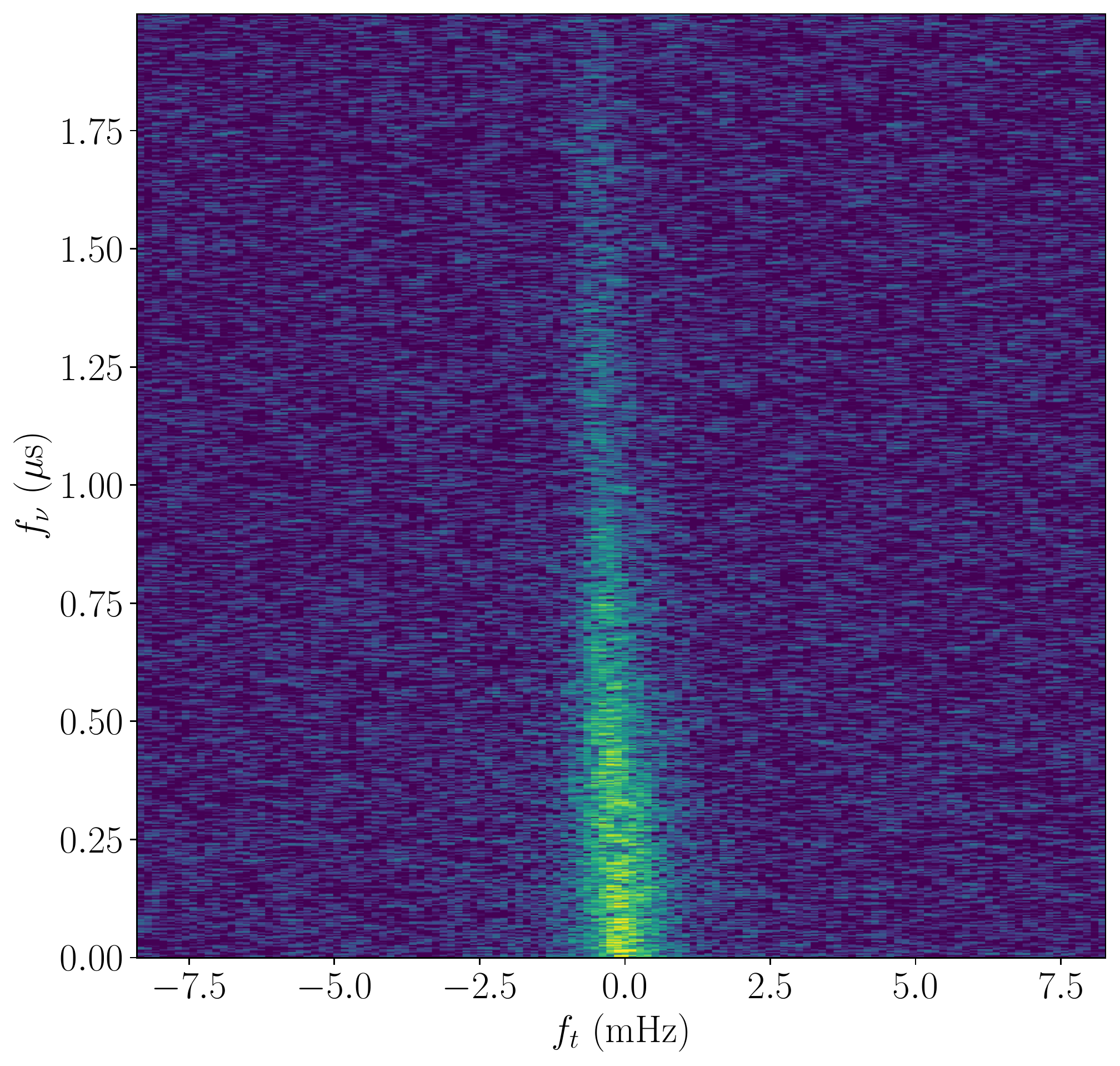}
\caption{Autocorrelation function (left) and secondary spectrum (right) from 2 June 2012 (MJD 55349) for PSR~J1603$-$7202. A strong phase gradient at this epoch presents as a tilted ACF and a highly asymmetric scintillation arc, which persists only for $\sim$days.}
\label{fig:compact}
\end{figure*}

At MJD 55350 we identify a short timescale ($\sim$days increase in the phase gradient, which coincides with a decrease in $\Delta\nu _d$ (marked with a grey dotted line in Figure \ref{fig:1603}). The ACF and secondary spectrum from one observation in this epoch is shown in Figure \ref{fig:compact}. The secondary spectrum demonstrates a clearly asymmetric power distribution in the scintillation arc, with a greater extent in differential delay (frequency Fourier conjugate, $f_\nu$) than most observations in this dataset \citep{Walker+22}. The decrease in $\Delta\nu _d$ (and the corresponding increase in scattering timescale) can be attributed to scattering from larger angles, which results in an increased scintillation strength. This is likely due to a dense and highly compact structure that would not be captured in the smooth DM variation models used in pulsar timing. The magnitude of both the phase gradient and bandwidth drop is only surpassed in this dataset by the known ESE. Phase gradient monitoring may be the only method for estimating the ${\rm DM}(t)$ induced by such structures for many pulsars.

Following the sharp decrease in $\Delta{\rm DM}_{\phi}(t)$ marked in the right panels of Figure \ref{fig:1603} (also discussed from Figure \ref{fig:J1603_1}), we observe that the variance in $\nabla\phi_\parallel$ decreases and $\Delta\nu_d$ increases to levels not observed since before the ESE. This further supports the idea that the scattering screen that dominated for the previous $\sim 2000\,$days has decreased in density in this time. Towards the end of our dataset, the variance in $\nabla\phi_\parallel$ begins to increase again and $\Delta\nu_d$ decreases, so the density minimum centred near MJD 56750 may have been a temporary ``hole" in the dominant scattering screen.

The $\Delta{\rm DM}_{\phi}(t)$ clearly shows a significant increase during the ESE, but the reconstruction of its shape (mainly width) differs from the timing-derived $\Delta{\rm DM}(t)$. The difference is significant, as shown by the 100 simulations of $\Delta{\rm DM}_{\phi}(t)$, and likely highlights a persistent anisotropy or weakness of our $\mathbf{V}_{\rm eff}$ model. In addition, since we do not have infinite sampling, the exact inflection and turning points for the ESE curve are not captured exactly, which can lead to further errors in the reconstruction. For this reason the ESE is perhaps best identified by the large $\nabla\phi_\parallel$ values.

\section{Discussion}
\label{sec:discussion}
Large gradients in DM$(t)$ are expected when discrete structures like ESEs, rather than Kolmogorov turbulence, cross the LOS. These structures are compact and dense, and as a result can easily dominate the scattering. In this case, we can probe these structures using scintillation and should expect to see large phase gradients. We have shown that the phase gradient parallel to velocity, $\nabla\phi_\parallel$ is a measure of the derivative of the transverse component of DM$(t)$, and can therefore reveal when large changes in DM are occurring. The precision to which it can predict $\Delta$DM$(t)$ depends on the stability of the IISM, the cadence of observations, whether the $\mathbf{V}_{\rm eff}$ is known, and the length of the prediction. The technique works best for short-timescale DM events since integrating the measurements of $\nabla\phi_\parallel$ produces a correlated noise process - increasing the uncertainty with time. However, this technique may remain useful for identifying ESEs, and for separating genuine DM$(t)$ variations from apparent variations due to chromatic pulse profile shape changes.

Our method will likely be more useful for detecting the ESEs in the phase gradients directly (e.g. using the middle panels of Figure \ref{fig:1603}), before fitting a velocity model and integrating the measurements to estimate its form. The phase gradient measurements are expected to have a powerlaw form and could therefore also be filtered (e.g. using a Wiener filter) to reduce the effect of the white measurement noise. The reconstruction will be more accurate for solitary pulsars and binaries with multiple observations across an orbit.

If pulsar timing and/or scintillation studies can solve for $s$ and $V_{\rm eff}$, this technique may also be used to constrain the pulsar distance, $D$ as the only unknown in Equation \ref{eqn:tpk}. In addition, the difference in gradient between the $\Delta {\rm DM}(t)$ measured directly with pulsar timing, and that inferred from phase gradients, may in part be due to the pulsar's radial velocity. Under this assumption, we can generate constraints on the radial velocity and the IISM density local to the pulsar.

We were able to show that the known ESE in PSR~J1603$-$7202 is apparent in the phase gradient measurements, demonstrating that the gradients measured in short $\sim 1\,$hr observations persist for many months. Because of this, phase gradient measurements with low duty cycle can recover DM$(t)$ on larger time scales, which we also demonstrated using a simulation. As the observing cadence decreases, so to does the accuracy of the recovered $\Delta {\rm DM}_\phi (t)$ measurements. Indeed we were able to identify a brief epoch of increased scintillation strength with a strong phase gradient that resembled a small ESE, but the observations were not dense enough across this time to recover a DM$(t)$ waveform.

During pulsar observations, the flux may be monitored in real-time to assess whether the pulsar is ``scintillating up" (is in a bright state), or not. This monitoring is a way to optimise pulsar timing array observing time, by focusing on pulsars at their brightest to achieve precise time of arrival measurements. If the phase gradients are also monitored in real-time, we may be able to determine when large DM gradients are occurring, and accordingly increase observing cadence to capture the variations. Without this, imperfect DM modelling in pulsar timing datasets will contribute excess low-frequency noise that impacts sensitivity to gravitational waves. This is particularly true for DM models that assume smooth variations (like linear interpolation across a wide time window, or Gaussian processes), as these will not capture the $\Delta {\rm DM}(t)$ induced by discrete compact structures.

As with PSR~J1603$-$7202, it is difficult to produce a complete model of $\mathbf{V}_{\rm eff}$ for most binary pulsars, so the exact values of $\Delta {\rm DM}_\phi$ are unlikely to be useful for correcting precision pulsar timing observations. For relativistic binaries, like the double pulsar J0737$-$3039A \citep{Kramer+21}, a full-orbit is often captured in a single observation. In this case, $\nabla\phi_\parallel$ can be measured at multiple points in the orbit, and the gradient in the direction of the mean $\mathbf{V}_{\rm eff}$ could be determined.

Another application may be studying the IISM density fluctuations at smaller spatial scales (or higher frequencies) than the DM$(t)$ usually allows, which may reveal evidence for a turbulence inner scale. Using $\nabla\phi$ to estimate the turbulence spectrum is worthwhile because the gradient reduces the spectral exponent by 2, and therefore is almost whitened and will not show spectral leakage.

\section{Conclusions}
\label{sec:conclusions}

We have shown that phase gradients measured through scintillation can be used to measure variations in electron density within a scattering screen over short but useful timescales. This was demonstrated both using a simulation, and by recovering a known ESE in the millisecond pulsar J1603$-$7202. We also described the detection of a compact event notable for its large phase gradient, which may be attributed to a small ESE-like structure. Our observation of sudden negative DM gradients marked the end of a $\sim$2000\,day period of enhanced scattering following the main ESE. The phase gradients may therefore detect ESEs and other large gradients in DM. They can potentially be used to separate true DM variations from changes to intrinsic pulse profile evolution with frequency \citep{Shannon+16}, if such changes are rapid \citep[e.g.][]{Lam+18}. Rapid DM changes must originate from compact structures, and should therefore also dominate the scattering.

The technique requires the detection of chromatic aberration in the scintillation pattern, which results from interstellar scattering. As a result, the recovered DM variations only describe the density variations in the scattering medium itself. In cases where these variations dominate the line of sight, such as in PSR~J1603$-$7202, the DM variations may also correspond to the total DM$(t)$ measured with pulsar timing. We inferred the phase gradients in the direction of $\mathbf{V}_{\rm eff}$ from a sheer in the ACF, but since the secondary spectrum contains the same information, the technique can also be applied to phase gradients inferred from the apex location and asymmetric power distribution of a scintillation arc \citep{Cordes+06}.

The technique will be especially valuable for the hundreds of scintillating pulsars with timing precision that is too low to measure DM variations directly. However, because the recovered $\Delta {\rm DM}_\phi$ depends on knowledge of $\mathbf{V}_{\rm eff}$, the result will be most accurate for solitary pulsars, or binaries with orbital periods longer than the observing cadence. Long-term monitoring of the scintillation properties will help to improve models of $\mathbf{V}_{\rm eff}$, through detection of annual and/or orbital variations in scintillation timescales or arc curvatures.

\section*{Acknowledgements}

The observations of PSR~J1603$-$7202 were collected as part of the Parkes Pulsar Timing Array program. Murriyang, the Parkes 64\,m radio telescope is part of the Australia Telescope National Facility (\url{https://ror.org/05qajvd42}), which is funded by the Australian Government for operation as a National Facility managed by CSIRO. We acknowledge the Wiradjuri people as the Traditional Owners of the Observatory site. This research was funded partially by the Australian Government through the Australian Research Council (ARC), grant CE170100004 (OzGrav).

\section*{Data Availability}

The version of the data analysis package (\textsc{scintools}) used at the time of this publication is preserved with the ``pre-release 0.3" tag. The dynamic spectra for PSR~J1603$-$7202 were published previously by \citet{Walker+22}, with a CSIRO Data Access Portal (DAP) address within. The simulation and other materials are available upon reasonable request to the corresponding author.



\bibliographystyle{mnras}
\bibliography{references.bib} 







\bsp	
\label{lastpage}
\end{document}